\documentclass[twocolumn,showpacs,preprintnumbers,amsmath,amssymb]{revtex4}
\usepackage{graphicx}% Include figure files
\usepackage{dcolumn}% Align table columns on decimal point
\usepackage{bm}% bold math
  \usepackage{latexsym,bm,amsmath,amssymb,amsfonts}
    \usepackage{makeidx}
\newcommand{\beq}{\vspace{-.4cm}\begin{eqnarray}}
\newcommand{\eeq}{\vspace{-.5cm}\end{eqnarray}}

\begin{document}

%\preprint{APS/123-QED}

\title{Thermal hadron spectrum in $e^+e^-$ annihilation from gauge/string duality}% Force line breaks with \\

\author{Yoshitaka Hatta}
\author{Toshihiro Matsuo}%
\affiliation{ Graduate School of Pure and Applied Sciences, University
of Tsukuba, \\Tsukuba, Ibaraki 305-8571, Japan
}%

\date{\today}% It is always \today, today,

\begin{abstract}
We compute the inclusive spectrum  of produced particles in $e^+e^-$ annihilation in confining gauge theories that have a
gravity dual
 and show that the momentum distribution exhibits the thermal behavior.
\end{abstract}

\maketitle
\emph{Introduction}---One of the puzzling features of hadron production in collider experiments is that the
inclusive rates are  well described by assuming the  `Boltzmann' distribution at low momenta
\begin{align}
\frac{dN_i}{d^3 p}\sim \exp\left(-\sqrt{p^2+m^2_i}/T\right)\,, \label{1}
\end{align}
where the subscript $i$ labels different hadron species. The parameter $T$, often referred to as the `temperature', is more or less independent of the collision
energy and
is typically of the order of the confinement scale $\gtrsim 150\,$MeV. The model (\ref{1}) gives a good fit of the identified particle yields in
 $e^+e^-$ annihilation \cite{Becattini:1995if,Becattini:2008tx,Chliapnikov:1999qi} (see, however, \cite{Andronic:2008ev}), hadron collision \cite{Becattini:1997rv} and heavy--ion collision experiments \cite{Becattini:2000jw,BraunMunzinger:2001ip}. Despite these phenomenological successes, however, the origin of the exponential behavior is not understood. One may invoke Hagedorn's picture of particle production  \cite{Hagedorn:1965st} in which an equilibrated state of secondaries (`fireball') is instantly formed after the collision.  Although such a scenario may sound plausible in the heavy--ion case where one expects the formation of the quark--gluon plasma, the situation is highly obscure in the case of $e^+e^-$ or  hadron collisions where the produced particles  have essentially no chance to interact among themselves.

In this letter we derive the exponential
distribution in $e^+e^-$ annihilation from gauge/string duality, thereby suggesting that
   the apparent thermal behavior is a generic feature of the strong coupling dynamics of gauge theories in the nonperturbative regime. [See \cite{Evans:2007sf} for a quite different approach to particle production using gauge/string duality. See also \cite{Kharzeev:2005iz,Castorina:2007eb,Bialas:1999zg,Hormuzdiar:2000vq} for earlier speculations concerning the behavior (\ref{1}).]
%Because of this difficulty,
%several authors have suggested that  (\ref{1}) could be a generic property of hadronization, or the %QCD vacuum itself, rather than something related to the actual temperature  .
 $e^+e^-$ annihilation in ${\mathcal N}=4$ supersymmetric Yang--Mills (SYM) theory and its variants has been recently analyzed in the framework of the AdS/CFT correspondence \cite{Maldacena:1997re}.
Here we briefly summarize the key features  at large 't Hooft coupling $\lambda \gg 1$.
(i) Unlike in QCD and in ${\mathcal N}=4$ SYM at weak coupling, there are no jets in the final state \cite{Hofman:2008ar,Hatta:2008tx,Hatta:2008tn}. The distribution of  energy is  spherical \cite{Hofman:2008ar}.
(ii) The average multiplicity scales almost linearly with the virtual photon energy $Q$ \cite{Hatta:2008tx,Hatta:2008tn}: $n(Q)\propto (Q/\Lambda)^{1-3/2\sqrt{\lambda}}$  where $\Lambda$ is an infrared cutoff. (iii)
The inclusive spectrum is peaked at the kinematical lower limit. More precisely, it has the form
\cite{Hatta:2008tn}
\begin{align}
\frac{dN}{dx}=\frac{Q^2}{\Lambda^2} F\left(\frac{Q}{\Lambda}x\right)\,, \label{we}
\end{align}
where $x=2E/Q$ is the Feynman variable.  Since the distribution is spherical, this is equivalent to
 \begin{align}  2E\frac{dN}{d^3p}=\frac{Q}{\pi p \Lambda^2}F\left(\frac{Q}{\Lambda}x\right)\,. \label{fin}
\end{align}
The function $F(y)$ remained unknown except for the property that it decays faster than the power--law when $y\gg 1$.
In \cite{Hatta:2008tn}, it was conjectured that
\begin{align} F\left(\frac{Q}{\Lambda}x\right) \propto \exp\left[  -c\left(\frac{Q}{\Lambda}x\right)^a\right]=  \exp\left[  -c\left(\frac{2E}{\Lambda} \right)^a\right]\label{tun}
\end{align}
where $c$ is a number of order unity. We shall show that a calculation
based on gauge/string duality yields the value $a=1$ in accordance with the phenomenological distribution (\ref{1})  with $1/T=2c/\Lambda$. \\

% Before starting the discussion, we wish to recall  that the weak coupling methods have been quite successful in %describing many aspects of jet physics in $e^+e^-$ annihilation \cite{Dokshitzer:1991wu}.
 %In particular, the observed `hump--backed' shape for the inclusive distribution of charged particles %\cite{Boehrer:1996pr} is consistent with the perturbative QCD prediction taking into account the gluon coherence %\cite{Dokshitzer:1991wu}.
% On the other hand, for genuinely nonperturbative quantities such as the relative yields of hadrons, statistical models %with only a few adjustable parameters do seem to
% capture the observed regularity in the data.
%  It is this aspect of particle production we expect that methods based on gauge/string duality might be a good starting %point for theoretical understanding.

\emph{Statistical model}---As a preliminary, let us first  point out the possible relationship  between particle production in strongly coupled gauge theories
  and a statistical model of $e^+e^-$ annihilation proposed by Bjorken and Brodsky long ago \cite{Bjorken:1969wi}.  Suppose, as is the case in QCD, that the multiplicity is dominated by the lightest particles (`pions') with mass $m$. The  inclusive distribution is given by
\begin{align} & 2E\frac{d\sigma }{d^3p}=\frac{e^4}{2 (2\pi)^3 Q^6} (k_\mu k'_\nu+k_\nu k'_\mu-g_{\mu\nu}k\cdot k') \\ & \times \sum_X
\langle 0|j^\mu(0)|p,X \rangle \langle p,X|j^\nu(0) |0\rangle
 (2\pi)^4 \delta^{(4)}(q-p-p_X)\,, \nonumber  \end{align}  where $k$ and $k'$ and electron and positron momenta, respectively and
 $j_\mu$ is a component of the $\mathcal{R}$--current operator which couples to the external virtual photon. Rather than directly dealing with this expression,  following  \cite{Bjorken:1969wi} we consider  the
cross section of producing exactly  $n$ pions and relate it to the inclusive distribution.
%[We discuss the nature of the single particle states $|p_i\rangle$ later.]
\begin{align} &\sigma_n =\frac{e^4}{2Q^6}(k_\mu k'_\nu+k_\nu k'_\mu-g_{\mu\nu}k\cdot k')\prod_{i=1}^n  \int \frac{d^3p_i}{2E_i(2\pi)^3} \nonumber \\
&\times \langle 0|j^\mu(0)|p_1,...p_n\rangle \langle p_1,...p_n|j^\nu(0) |0\rangle
 (2\pi)^4 \delta^{(4)}(q-\sum_i p_i)\,.  \nonumber \end{align} We take $n$ to be large, of the order of the average multiplicity $n \sim Q/\Lambda \gg 1$.
 The model assumes that the  hadronic matrix element  can be written  as
\begin{align}  & \langle 0|j^\mu(0)|p_1,...p_n\rangle \langle p_1,...p_n|j^\nu(0) |0\rangle \nonumber \\
& \to a_n (q^\mu q^\nu-g^{\mu\nu}q^2)\, e^{-\beta Q} \nonumber \\
&\to  a_n (q^\mu q^\nu-g^{\mu\nu}q^2) e^{-\beta (E_1+\cdots +E_n)}\,, \label{cc} \end{align} where $\beta= 1/T$ and the arrow means that the replacement is valid only  under the $n$--particle phase space integral. Note that the matrix element itself gives the `thermal' factor, which distinguishes this model from other statistical models where the factor
comes from the total phase space. [See also, \cite{Engels:1973pf}.]
We then get
\begin{align}
\sigma_{tot}=\sum_n \sigma_n &= \sum_n \frac{e^4 a_n}{2Q^2 }\prod_i^n   \int \frac{d^3p_i}{2E_i(2\pi)^3}  e^{-\beta(E_1+\cdots +E_n)} \nonumber \\ & \qquad \times (2\pi)^4 \delta^{(4)}(q-\sum_i^n p_i)\,. \label{to} \end{align}
Due to the non--renormalization theorem for the $\mathcal{R}$--current correlator \cite{Anselmi:1997am}, the total cross section is given by the one--loop result
for all values of the coupling
 \begin{align}
\sigma_{tot}=\frac{e^4N_c^2}{32\pi Q^2}\,. \end{align}
Therefore,
\begin{align} \frac{2E}{\sigma_{tot}}\frac{d\sigma}{d^3p}=&\sum_n \frac{16\pi a_n}{N_c^2  } n e^{-\beta E} \prod_i^{n-1}  \left( \int \frac{d^3p_i}{2E_i(2\pi)^3}  e^{-\beta E_i} \right) \nonumber \\ & \qquad \qquad \qquad \times 2\pi \delta^{(4)}(q-\sum_i^n p_i)\,. \label{take} \end{align}
After an integration over $Q$ within an interval $\Delta Q \ll Q$, (\ref{take}) takes the form
\begin{align} &\Delta Q\, 2E\frac{dN}{d^3p}  =\sum_n \frac{32\pi^2 a_n}{N_c^2 } n e^{-\beta E} \nonumber \\
 & \qquad \times \int \frac{d^3x}{(2\pi)^3} e^{i\bf{p}\cdot\bf{x}} \left(  \int \frac{d^3p'}{2E'(2\pi)^3}  e^{-\beta E'+i\bf{p'\cdot x}} \right)^{n-1} \nonumber \\
% &\approx \sum_n \frac{4 a_n}{\pi N_c^2  } n e^{-\beta E} \int d^3x e^{ipx}
%\left(\frac{1}{4\pi^2(\beta^2+x^2)}\right)^{n-1} \nonumber \\
%& \approx \sum_n \frac{4 a_n}{\pi N_c^2  } n e^{-\beta E} \int d^3x e^{ipx} \left(\frac{1}{4\pi^2 \beta^2}\right)^{n-1} %e^{-(n-1)\frac{x^2}{\beta^2}} \nonumber \\
&
\approx \sum_n \frac{4 a_n}{\pi N_c^2  } n e^{-\beta E} \left(\frac{1}{4\pi^2 \beta^2}\right)^{n-1} \left(\frac{\pi \beta^2}{n-1}\right)^{3/2}e^{-\frac{p^2\beta^2}{4(n-1)}}\,, \nonumber
\end{align} where we approximated $E'\approx p'$. The last factor may be omitted and the summation simply gives a factor $\Delta n \sim \Delta Q/\Lambda$
\begin{align}  2E\frac{dN}{d^3p} \sim \frac{4 a_n}{\pi N_c^2 \Lambda } n e^{-\beta E} \left(\frac{1}{4\pi^2 \beta^2}\right)^{n-1} \left(\frac{\pi \beta^2}{n}\right)^{3/2}\,.  \label{key} \end{align}
The coefficients $\{a_n\}$ can be determined by  matching with the total cross section (\ref{to})  \cite{Bjorken:1969wi}
\begin{align} a_n&=(n-1)!(n-2)!\frac{N_c^2 (4\pi \beta)^{2n-4}}{(2n-4)!} \nonumber \\
% &\approx \frac{n!n!(2n)^4N_c^2 (4\pi \beta)^{2n-4}}{nn^2 (2n)!} \nonumber \\
&\approx n^{3/2}\sqrt{\pi} N_c^2
(2\pi \beta)^{2n-4}\,. \label{comt}\end{align}
 [We have used  Stirling's formula $n! \approx \sqrt{2\pi}n^{n+1/2}e^{-n}$ valid for $n\gg 1$.]
Inserting this into (\ref{key}), we obtain
\begin{align}  2E\frac{dN}{d^3p}
\sim \frac{  n \beta}{\pi \Lambda }  e^{-\beta E} \sim \frac{Q\beta}{\pi \Lambda^2} e^{-\beta E}\,. \end{align}
 This agrees with (\ref{fin}) provided $a=1$ and
 \begin{align}
 F(y)\propto \sqrt{y^2-\frac{4m^2}{\Lambda^2}}e^{-cy}\,.
 \end{align}
 The proportionality constant is determined from the energy conservation  \begin{align} 2=\int^1_{2m/Q} x \frac{dN}{dx} dx\,. \end{align}
  The result is
 \begin{align} \frac{dN}{dx}=\frac{c}{2K_2\left(\frac{2mc}{\Lambda}\right)} \frac{Q^3}{m^2\Lambda}\sqrt{x^2-\left(\frac{2m}{Q}\right)^2}
 e^{-cQx/\Lambda}\,, \end{align} and for the average multiplicity,
 \begin{align} \langle n\rangle =\frac{Q}{m}\frac{K_1\left(\frac{2mc}{\Lambda}\right)}{K_2\left(\frac{2mc}{\Lambda}\right)}\,. \end{align}
  It is straightforward to include heavier hadrons with mass $m^*>m$ as a small contamination in the final state. Proceeding as before, we find \begin{align}  & 2E^*\frac{dN^*}{d^3p}=\frac{4 a_n}{\pi N_c^2 \Lambda } \sum_{l=0}^{n} {}_n\mathrm{C}_l\,  l\, e^{-\beta E^*} \int \frac{d^3x}{(2\pi)^3} e^{i\bf{p\cdot x}}
  \nonumber \\ & \times \left( \int \frac{d^3p'}{(2\pi)^3}  \frac{e^{-\beta E'+i\bf{p'\cdot x}}}{2E'} \right)^{n-l} \left( \int \frac{d^3p'}{(2\pi)^3}  \frac{e^{-\beta E'^*+i\bf{p'\cdot x}} }{2E'^*}\right)^{l-1} \nonumber \\
 & =  \frac{4 a_n}{\pi N_c^2 \Lambda } n e^{-\beta E^*} \int \frac{d^3x}{(2\pi)^3} e^{i\bf{p\cdot x}}
  \nonumber \\ & \times \left( \int \frac{d^3p'}{(2\pi)^3}  \frac{e^{-\beta E'+i\bf{p'\cdot x}}}{2E'} + \int \frac{d^3p'}{(2\pi)^3}  \frac{e^{-\beta E'^*+i\bf{p'\cdot x}}}{2E'^*} \right)^{n-1} \nonumber \\
 &  \approx e^{-\beta(E^*-E)} 2E\frac{dN}{d^3p}\,.\end{align} [Since typically $l\ll n$, we assume that the $l$--dependence of $a_n$ is negligible, or trivially factorized in the form $a_{n,l}\sim a_n b^{l}$ with $b$ being a number of order unity.] This immediately leads to the characteristic ratio \begin{align}
 \frac{\langle n^*\rangle }{\langle n\rangle}= \frac{m^*K_1(\beta m^*)}{mK_1(\beta m)} \sim e^{-\beta(m^*-m)}  \label{ther}\end{align}
  which resembles the observed scaling \cite{Becattini:1995if,Becattini:2008tx,Chliapnikov:1999qi,Becattini:1997rv,Becattini:2000jw,BraunMunzinger:2001ip}.
Therefore, finding the exponential factor $e^{-\beta Q}$ in  (\ref{cc}) establishes a connection between strongly coupled ${\mathcal N}=4$ SYM and a statistical model of particle production in $e^+e^-$ annihilation.\\

\emph{Gauge/string duality}---Let us now turn to  the matrix element
\begin{align} \langle 0|\epsilon \cdot j(0)|p_1,...p_n\rangle\,, \label{mat} \end{align} where $\epsilon_\mu$ is an arbitrary polarization vector which we take to be transverse ($\epsilon \cdot q=0$) for convenience.
 The prescription
  for evaluating amplitudes like (\ref{mat}) has been spelled out in \cite{Polchinski:2001tt,Polchinski:2002jw}.
  In the limit $\lambda \to \infty$, this can be viewed as the scattering of $n+1$ bulk fields in $\mbox{AdS}_5\times S^5$ in the supergravity approximation.
We shall use the Poincar\'e coordinates
\begin{align} ds^2=R^2 \frac{-dt^2+d\vec{x}^2+dz^2}{z^2}+R^2 d\Omega_5^2\,. \end{align} The Minkowski boundary is at $z=0$ and
  the space is cut off at $z=1/\Lambda$.
  The current $j^\mu$ is dual to a five--dimensional Kaluza--Klein photon whose wavefunction solves the Maxwell equation in the bulk. For timelike momenta, $q^2=Q^2>0$, and with the infalling--wave condition as appropriate for the problem of jet evolution \cite{Hatta:2008tx}, it features the Hankel function
 \begin{align} A_\mu(x^\mu,z) = \epsilon_\mu e^{-iqx}\frac{\pi N_c  Qz}{8\pi^{5/2}R^3} H_1^{(1)}(Qz)\,.
 %  \nonumber \\
 %A_z(x,z)=iq\cdot \epsilon \,e^{-iqx}\frac{\pi N_c z}{8\pi^{5/2}R^3}H_0^{(1)}(Qz)\,.
 \label{pho}\end{align}
 [The normalization is taken from \cite{Hong:2004sa}. Note that $A_z=0$ in the current gauge $(\epsilon \cdot q=0)$.]
 Alternatively,  (\ref{pho}) may also be regarded as the wavefunction of a heavy vector meson with which the photon couples in a way similar to the vector meson dominance in QCD.
  Then the factor $N_c$ in (\ref{pho}) acquires the meaning as the strength of this coupling. Naturally, the timelike photon or the meson decays, and at strong coupling
   the decay is so complete that the final state contains only particles with the smallest virtuality $p_i^2= m^2 \sim \Lambda^2$ \cite{Hatta:2008tx,Hatta:2008tn}.
 These are represented, for simplicity, by the scalar wavefunction in $\mbox{AdS}_5$. We shall consider two scenarios.
% \begin{align} e^{ip_ix} z^2J_{\Delta_+-2}(mz), \qquad e^{ip_i x}z^2N_{\Delta_+-2}(m z)\,,  \end{align} where $\Delta_+$ is the conformal dimension of the operator which creates the state $|p\rangle$.
 %The former is a normalizable mode while the latter is a non--normalizable mode for $\Delta_+\ge 3$. We shall consider two %scenarios.
 (i) Hadrons corresponding to normalizable modes in the cutoff AdS space. They have the wavefunction
 \begin{align} \Phi_i(x,z)=e^{ip_ix} \frac{\sqrt{2} \Lambda z^2}{2\pi^{3/2}R^4} J_{\Delta_+-2}(m z)\,,\label{part} \end{align} where  $\Delta_+$ is the conformal dimension of the interpolating operator.
 The lightest one is dual to the lowest dimension scalar operator in the ${\mathcal N}=4$ graviton supermultiplet. This has $\Delta_+=2$ and
    $m\approx 2.4\Lambda$ as determined from the first zero of the Bessel function $J_0$.
 (ii) The non--normalizable mode with $\Delta_-=4-\Delta_+=1$
  \begin{align} \Phi_i(x,z)=e^{ip_ix} \frac{\pi i \Lambda z^2}{2\pi^{3/2}R^4} H_1^{(1)}(\Lambda z)\,. \label{part} \end{align}
 This somewhat peculiar choice is motivated by the following argument. As noted in \cite{BallonBayona:2007rs} and emphasized in \cite{Pire:2008zf}, the value $\Delta_-=1$ realizes Bjorken scaling in DIS which is the hallmark of the parton picture.
[See, also, \cite{Hatta:2007he}.]
 Curiously, these modes contain the so--called singletons, or rather, doubletons
  \cite{Gunaydin:1984fk}
   that are often referred to as the most elementary
excitations, or `partons' in AdS spaces. Moreover, the bulk supergravity modes such as (\ref{pho}) are bound states of singletons \cite{Ferrara:1998jm,Ferrara:1998ej}. This fits nicely with the branching picture at strong coupling proposed in \cite{Hatta:2008tx}. Note that at the borderline value $\Delta_-=1$ the normalization integral suffers a logarithmic UV ($z\to 0$) divergence. This is analogous to the fact that the light--cone wavefunction with soft partons (gluons) in QCD is not normalizable due to the same reason \cite{Mueller:1993rr}.
  %(see, e.g., \cite{Hatta:2007he}).

%  Instead of scalars we may also use their ${\mathcal N}=4$ supersymmetric partners (fermions and gauge bosons).
% Then the twist $\tau=\Delta_- -j$ appears in place of $\Delta_-$. For the singletons, $\tau$ is always 1 ($3/2-1/2=1$ for %fermions, $2-1=1$ for gauge bosons) so the wavefunction
%is unchanged up to the spin structure.

In the branching picture of \cite{Hatta:2008tx}, one evaluates diagrams with only the three--point ($1 \to 2$) vertices. On the other hand, the supergravity effective action contains all possible higher--point vertices allowed by symmetry. To better illustrate the essential point of our calculation, let us first consider a contribution to the matrix element from a local $(n+1)$--point vertex of $n$ scalars and a photon.
The zero mode integration reads, symbolically,
\begin{align} &\langle 0|\epsilon \cdot j(0)|p_1,...p_n\rangle \nonumber \\ & \sim \frac{g_c^{n+1}}{\alpha' g_c^2}
\int dz d\Omega_5 \sqrt{-G} F(\alpha' \partial^2 )(\Phi)^n A_\mu \,, \label{so} \end{align} where  $g_c\sim g_s\alpha'^2$ is the closed string coupling.
 The simplest vertex is obtained by expanding the function $F$ to linear order in its argument
 \begin{align} & \frac{g_c^{n+1}}{\alpha' g_c^2}
\int dz d\Omega_5 \sqrt{-G}  \frac{ \Phi^{n-2}\alpha' v^a\partial_a \Phi \partial_\mu \Phi }{R}G^{\mu\nu} A_\nu
   \nonumber \\ &\sim  g_s^{n-1}\alpha'^{2n-2} N_c \int dz d\Omega_5 \frac{R^{10}}{z^5}    \sum_{ij}
   ({\mathcal Q}_i p_j+{\mathcal Q}_j p_i) \cdot \epsilon  \nonumber \\ &
     \quad \times \left(\frac{\Lambda z^2}{R^4}\left\{\begin{array}{c}
                                           J_0(mz) \\
                                           H_1^{(1)}(\Lambda z)
                                         \end{array} \right\}
  \right)^n  \frac{ Qz^3}{R^6} H_{1}^{(1)}(Qz)\,. \end{align} where $v^a$ ($a=1,2,..,5$) is a Killing vector on $S^5$ and ${\mathcal Q}_{i,j}$ are the U(1) charges. The sum is over oppositely charged pairs.
 Since $n\sim Q/\Lambda$ is large, one can do the saddle point approximation.
 It is consistent to look for a saddle point in the asymptotic region of the Hankel function $H_{1}^{(1)}(Qz) \sim e^{iQz}/\sqrt{Qz}$.
The equation determining the saddle point is, setting $n=kQ/m$ where $k$ is a number of order unity,
\begin{align} \label{uu} -iQ= \frac{kQ}{m}\frac{d}{dz}\ln\left(z^2 J_0(m z)
\right)\,,
 \end{align} and  similarly for  $H_1^{(1)}(\Lambda z)$ in which case we set $n=kQ/\Lambda$.
 We solve this equation  numerically in the form
  \begin{align} z=\frac{\alpha}{\Lambda}i\,, \nonumber \end{align} and find that, in the physically interesting range $0.5\lesssim k \lesssim 1.0$,  $\alpha \approx 0.2-(0.8\pm 0.1)i$ and $\alpha \approx 0.5\pm 0.1$ for the $J_0$ and $H_1^{(1)}$ cases, respectively. This leads to, using $4\pi g_s\alpha'^2/R^4=1/ N_c$ and ${\mathcal Q}_i=-{\mathcal Q}_j$,
\begin{align} &\langle 0|\epsilon \cdot j(0)|p_1,...p_n\rangle  \nonumber\\ & \sim
 \frac{N_c}{N_c^{n-1}} \,n \sum_i {\mathcal Q}_i p_i \cdot \epsilon  \left(\frac{1}{\Lambda}\right)^{n} \frac{Q}{z^2\sqrt{Q/z}}\frac{e^{-\alpha Q/\Lambda}}{\sqrt{Qz}}\Big\arrowvert_{z\sim i/\Lambda}\,, \nonumber
% \nonumber \\& \sim \frac{N_c}{N_c^{n-1}}\, n \sum_i {\mathcal Q}_i p_i \cdot \epsilon %\left(\frac{1}{\Lambda}\right)^{n-2} e^{-xQ/\Lambda}\, \nonumber
\end{align} where the factor $1/\sqrt{Q/z}$ is from the fluctuation around the  saddle point.
Squaring and noting that $p_i^\mu p_j^\nu \to g^{\mu\nu}\delta_{ij}p_i^2\sim g^{\mu\nu}\delta_{ij}\Lambda^2$ under the phase space integral, we find
\begin{align} |\langle 0| \epsilon \cdot j|p_1,...p_n\rangle |^2 \sim \frac{N_c^2}{N_c^{2n-2}} \frac{nQ^2}{\Lambda^{2n-4}} e^{-2cQ/\Lambda}\,, \label{last} \end{align} where $c=\mbox{Re}\, [\alpha]$. (\ref{last}) features the  exponential factor as in (\ref{cc}). Note that having $n\sim Q$ particles in the final state is crucial in order to obtain this behavior. If $n$ is order unity, the integral would be dominated by intermediate $z$ values, leading only to a power--law  \cite{Polchinski:2001tt}.

The evaluation of the nonlocal diagrams with  the 1 $\to$ 2 vertices is considerably more complicated, but without going into the details one can argue that they also give rise to the exponential factor. [The exact value of $c$ may be different, however.] Indeed, as one approaches the root of the branching tree, the intermediate states necessarily have an increasingly large number of constituents, hence are dual to operators with increasingly large dimensions whose wavefunctions are localized around $z\sim 1/\Lambda$. In particular, the photon couples to fields $\Phi_{X,Y}$ having dimensions of order $\Delta_{X,Y}\sim n\sim Q/\Lambda$ via the vertex  $A^{\mu}(\Phi_X\partial_\mu \Phi^*_Y-\Phi_Y\partial_\mu \Phi_X^*)$ [See \cite{BallonBayona:2007rs} for a related discussion including the issue of gauge invariance of this coupling.] The matrix element will be given by a multiple convolution of the form
 \begin{align}
(p_X^\mu-p_Y^\mu)\int dzdz'dz''\cdots \sqrt{-G}A_\mu(z) G_X(z,z')G_Y(z,z'')\nonumber \\
 \cdots G(z''',z_1)\Phi(z_1)\Phi(z_1)\cdots G(z'''',z_{n/2})\Phi(z_{n/2})\Phi(z_{n/2})\,,\nonumber  \end{align}
 where $G$'s are   propagators in (cutoff) $\mbox{AdS}_5$. At small $z$,  $G_{X,Y}(z,z')\sim z^{\Delta_{X,Y}}\sim z^{n}$, so the function obtained after the integration $dz'dz''\cdots$ is as strongly peaked around $z\sim 1/\Lambda$ as  $\Phi^n(z)$ in (\ref{so}). The final integral over $z$ then leads to the exponential factor by the same mechanism as in the local case above.

In fact, although the two contributions--local and nonlocal diagrams--have  similar $Q$--dependence, they crucially differ in $N_c$--dependence. When comparing (\ref{cc}) and (\ref{comt}) with (\ref{last}), one notices that the latter is enormously suppressed by the factor $1/N_c^{2n-2}$ ($1/N_c^{n-1}$ in the amplitude), which actually was expected from the standard large--$N_c$ counting.
  If the non--local diagrams also received this suppression, one would have to conclude that
 the multi--particle production process would be a negligible fraction of the total cross section (\ref{to}) which scales as $N_c^2$. Instead, the production cross section of a \emph{single} heavy vector meson, being ${\mathcal O} (N_c^2)$, would alone seem to saturate the total cross section. In order to avoid this unphysical conclusion,  the subsequent decay of the meson into $n$ particles via 1 $\to$ 2 splittings must somehow be an  ${\mathcal O}(N_c^0)$ effect rather than suppressed by a large negative power of $N_c$.  Fortunately, a similar problem was encountered and resolved in the context of $e^+e^-$ annihilation in (two--dimensional) QCD at large $N_c$  \cite{Einhorn:1976ev}. There it was observed that a proper inclusion of the decay width ($\Gamma\sim 1/N_c^2$ in the present situation) in the intermediate propagators  together with a suitable averaging of the incident energy give just the right number of $N_c$ factors to cancel
 the unwanted $1/N_c$ suppressions from the three--point couplings $g\sim 1/N_c$. Namely,
  the squared propagator in the momentum space reads \begin{align} \Big\arrowvert \frac{g}{p^2-m^2+i\Gamma}\Big\arrowvert^2=\frac{\Gamma}{(p^2-m^2)^2+\Gamma^2}\frac{g^2}{\Gamma}\,. \label{eff} \end{align} The first factor on the right hand side is unit normalized with respect to the integration over the momentum and the second factor is ${\mathcal O}(N_c^0)$, so effectively (\ref{eff}) is ${\mathcal O}(N_c^0)$ contrary to the naive expectation ${\mathcal O}(1/N_c^2)$. Since this argument appears to be quite general,
 we expect that the nonlocal diagrams will not be $N_c$--suppressed unlike the local ones.

   Keeping this qualification in mind, we conclude  that the distribution obeys the thermal law (\ref{1}), and especially, (\ref{ther}), though clearly the parameter $T\sim \Lambda$  has nothing to do with the temperature. In view of the strikingly different pictures of the final state in QCD and in strongly coupled SYM as mentioned in the introduction, it is intriguing that the latter can explain this particular nonperturbative aspect of particle production.

\emph{Acknowledgments}---This work is supported, in part, by
Special Coordination Funds for Promoting Science and Technology of
the Ministry of Education, Culture, Sports, Science and
Technology, the Japanese Government.

\end{document}